# Low Timing Jitter Detector for Gigahertz Quantum Key Distribution

R.J. Collins, R.H. Hadfield, V. Fernandez, S.W. Nam and G.S. Buller

A superconducting single-photon detector based on a niobium nitride nanowire is demonstrated in an optical-fibre-based quantum key distribution test bed operating at a clock rate of 3.3 GHz and a transmission wavelength of 850 nm. The low jitter of the detector leads to significant reduction in the estimated quantum bit error rate and a resultant improvement in the secrecy efficiency compared to previous estimates made by use of silicon single-photon avalanche detectors.

*Introduction*: Quantum Key Distribution (QKD) [1], [2], [3] offers a verifiably secure method to share cryptographic keys between two remote parties (Alice and Bob). QKD allows Alice and Bob to detect the presence of an eavesdropper (Eve) and to agree upon a secure encryption key for communication by encoding information on the states of individual photons.

We have previously constructed a QKD test bed based on the B92 protocol [2] using highly attenuated pulses from vertical cavity surface emitting lasers (VCSELs) operating at a wavelength of 850 nm in conjunction with a quantum transmission channel composed of standard telecommunications optical fibre (single-mode at wavelengths of 1300/1550 nm) [4][5][6]. In standard telecommunications optical fibre, the 850 nm wavelength light propagates as two modes ($LP_{01}$ and $LP_{11}$), and due to mode dispersion these two modes travel at different velocities, causing a spreading of the optical pulses that can lead to an increase in the quantum bit error rate (QBER).  In our test bed, the second-order mode is suppressed by utilising mode control methods, allowing us to achieve clock rates of up to 3 GHz using silicon single photon avalanche diodes (Si SPADs) [6],[7].  In this letter we present estimates of QBER for a QKD test bed operating at a frequency of 3.3 GHz and using a low-jitter superconducting single-photon detectors (SSPDs) based on niobium nitride nanowires [8],[9].

The SSPDs used in this work were 100 nm wide niobium nitride meander wires, covering a 10 µm x 10 µm area with a fill factor of 50 %.  Each superconducting wire was biased close to its critical current; when a photon strikes the wire a resistive hotspot is formed, triggering a measurable voltage pulse.  The SSPD detectors were fibre-coupled with single-mode telecom fibre, packaged and installed in a cryogen-free refrigerator system, with an operating temperature of 3 K [10].  These detectors offer low jitter (68 ps full width at half maximum - FWHM) [11] and fast recovery times (below 10 ns), making them highly promising candidates for QKD at gigahertz clock rates.

*Experimental Test bed:* Figure 1 shows a schematic of the QKD test bed.  The layout is described in detail in [4], [5] and [6].  The mean photon number per pulse leaving Alice ($\mu$) is $0.1 \pm 6 \times 10^{-3}$.  Different quantum channel transmission distances are simulated with a programmable digital attenuator.  At 850 nm an attenuation of 2.2 dB corresponds to a distance of 1 km in fibre.  In this demonstration we used a pair of SSPDs in place of Si SPADs.  Amplified voltage pulses from each SSPD were recorded separately by a time-correlated single-photon counting (TCSPC) card [12] in a desktop PC that was operated in histogram mode.

The QBER in a QKD system is defined as the ratio of error bits to the total number of bits received. In QKD, all errors are attributed to the presence of an eavesdropper (Eve). With rising QBER an increasing proportion of received bits must be expended in privacy amplification to minimize Eve's share of the information. In our B92 QKD scheme the distillation efficiency of secret key (the ratio of the final secure bit rate to the raw detector events) is given by [13]:

$$\text{Secrecy Efficiency} \approx \left[1 + Q\log_2(Q) - 3.5Q - I_{AE}\left(1 - (1-Q)\log_2(1-Q) - 3.5Q\right)\right] \quad (1)$$

where $Q$ is the QBER and $I_{AE}$ =0.29 is the maximum information shared between Alice and Eve for the B92 protocol [14]. Although the B92 protocol is inherently more susceptible to the "intercept and resend" attack with a lossless channel, it was chosen for experimental simplicity. This test bed was used to characterise optical components such as detectors in order to assess their suitability in future implementations of more secure QKD protocols.

In the context of high bit rate polarisation-encoding QKD there are usually three contributions to the QBER:

$$QBER = QBER_{Opt} + QBER_{Det} + QBER_{Int} . \quad (2)$$

The first term, $QBER_{Opt}$, is due to the polarisation extinction ratio resulting mainly from passive optical components within the optical system. This is a measure of the optical quality of the setup and is independent of transmission distance. The second term, $QBER_{Det}$, arises from the detector dark counts. At high link loss, the number of received bits falls and the contribution of dark counts to the QBER increases. The third term is the intersymbol interference, where errors are caused by counts from a given clock period that leak into the adjacent period due to source jitter, detector jitter, and dispersion in the transmission medium. This term usually sets the minimum QBER in high clock rate QKD using SPADs with ~400 ps FWHM jitter [5]. In this demonstration we are able to reduce this source of error by employing a detector with significantly lower jitter (SSPD, 68 ps FWHM).

*Results:* Figure 2 shows the QBER and net bit rate (NBR – the projected distillation rate of secret key, given by multiplying equation (1) by the raw event rate at the detectors) for the test bed at a clock rate of 3.3 GHz by use of the SSPD receiver. With this low-jitter detector we were able to achieve a QBER below 1 % out to 20 km simulated transmission distance. As a consequence of the increased loss in the quantum channel of 2.2 dBkm$^{-1}$, the rate of the raw events at the detector decreases with distance, leading to a corresponding decrease in the NBR. The dominant error contribution in the 1-20km range is the polarisation contrast, and not the dark count rate. At 25 ± 0.11 km the estimated QBER is 3.6 ± 0.2 %, due to the rising influence of dark counts (below 10 Hz per SSPD channel). Although the dark count probability within the coincidence window is low, the greatly reduced rate of incident photons at the detector means that the QBER shows a significant increase. This overall behaviour of the QBER contrasts with previous results in this test bed for Si SPADs with 400 ps jitter clocked at 2 GHz, where intersymbol interference prevented the QBER from falling below 5 %, and QBER rises above 11 % - the limit for secure transmission according to Equation (1) - at 12 km [5]. The detection efficiency of the SSPD at a wavelength of 850 nm is ~5 %, compared to 40 % for the Si SPAD, which leads to a corresponding reduction in the photon detection rate at Bob. However, the significantly lower timing jitter of the SSPD results in very low

QBERs, allowing an improved secrecy efficiency compared to that achieved by use of a Si SPAD.  As a result, the maximum NBR (at short transmission distances) was similar for SSPDs used in a 3.3 GHz clocked system and for Si SPADs in a 2 GHz clocked system [15], but the SSPDs offer a clear improvement in overall transmission range.

*Conclusions:* We have demonstrated the impact of an SSPD detector in a high bit rate QKD test-bed at a clock frequency of 3.3 GHz and a transmission wavelength of 850 nm.  The maximum NBR achieved is comparable to that of the more widely used Si SPADs at 2 GHz [15], with greatly improved secrecy efficiency.  The low jitter of the SSPD allows low QBER to be achieved for significantly longer simulated transmission distances (up to 25 km).  We emphasise that there is considerable scope for further improvements in performance using this detector in the realm of high bit rate QKD.  The SSPD can be used at 1310 nm or 1550 nm wavelengths, where fibre transmission losses are significantly lower, allowing the system transmission range to be extended significantly [16].   The low jitter of the detector (68 ps FWHM) means that system clock rates can in principle be increased to ~10 GHz.  Furthermore, new designs incorporating SSPDs into optical cavities allow significant improvements in detection efficiency (up to 57 %) [17]. Both the increased clock rates and detection efficiencies afforded by the SSPDs will lead to dramatic improvements in NBR.

**References**

1  BENNETT, C.H., and BRASSARD, G.: 'Quantum cryptography: public key distribution and coin tossing', Proceedings of the IEEE conference on computers, systems & signal processing, Bangalore, India, 1984, pp 175-179

2  BENNETT, C.H.: 'Quantum cryptography using any two nonorthogonal states', Physical Review Letters, 1992, 68, (21), pp 3121-3124

3  GISIN, N., RIBORDY, G., TITTEL, W., ZBINDEN, H.: 'Quantum Cryptography', Reviews of Modern Physics, 2002, 74, (1), pp 145-195

4  GORDON, K.J., FERNANDEZ V., TOWNSEND, P.D., and BULLER, G.S.: 'A short wavelength gigahertz clocked fiber-optic quantum key distribution system', IEEE Journal of Quantum Electronics, 2004, 40, (7), pp 900-908

5  GORDON, K.J., FERNANDEZ, V., BULLER, G.S., RECH, I., COVA, S.D., and TOWNSEND, P.D.: 'Quantum key distribution system clocked at 2GHz', Optics Express, 2005, 13, (8), pp 3015-3020

6  FERNANDEZ, V., COLLINS, R.J., GORDON, K.J., TOWNSEND, P.D., and BULLER, G.S.: 'Passive optical network approach to gigahertz-clocked multi-user quantum key distribution', IEEE Journal of Quantum Electronics, 2007, 43, (2), pp 130-138

7  Perkin Elmer SPCM-AQR-12 Slik™ thick junction Si APDs were used previously. Data sheet at Perkin Elmer Optoelectronics Canada Ltd., Vaudreuil, Quebec, Canada, http://optoelectronics.perkinelmer.com. Note: use of a trade name is for technical clarity and does not imply endorsement by NIST.

8  GOL'TSMAN, G.N., OKUNEV, O., CHILKOV, G., LIPATOV, A., SEMENOV, A., SMIRNOV, K., VORONOV, B., and DZARDANOV, A.: 'Picosecond superconducting single-photon optical detector', Applied Physics Letters, 2001, 79, (6), pp 705-707

9  VEREVKIN, A., ZHANG, J., SOBOLEWSKI, R., LIPATOV, A., GKHUNEV, O., CHULKOVA, G., KORNEEV, A., SIMROV, K., GOL'TSMAN, G.N., and SEMENOV, A.: 'Detection efficiency of large-active are NbN single-photon superconducting detectors in the ultraviolet to near-infrared', Applied Physics Letters, 2002, 80, (25), pp 705-707,

10  HADFIELD, R.H., STEVENS, M.J., GRUBER, S.G., MILLER, A.J., SCHWALL, R.E., MIRIN, R.P., and NAM, S.W.: 'Single-photon source characterization with a superconducting single-photon detector', Optics Express, 2005, 13 (26), pp 10846-10853

11  STEVENS, M.J., HADFIELD, R.H., SCHWALL, R.E., NAM, S.W., MIRIN, R.P., and GUPTA, J.A.: 'Fast lifetime measurements of infrared emitters using a low-jitter superconducting single-photon detector', Applied Physics Letters, 2006, 89, (3), pp 031109-1-031109-3

12  Becker and Hickl SPC-600 time-correlated single-photon counting (TCSPC) card. Note: use of a trade name is for technical clarity and does not imply endorsement by NIST.

13  TANČEVSKI, L., SLUTSKY, B., RAO, R., and FAINMAN, S: 'Evaluation of the cost of error-correction protocol used in quantum cryptographic transmission', 1998, SPIE Proceedings, 3228, pp 322-330

14  CLARKE, R.B.M., CHEFLES, A., BARNETT, S.M., and RIIS, E.: 'Experimental demonstration of optimal unambiguous state discrimination', Physical Review A, 2001, 63, (4), pp 040305—1-040305—4

15  FERNANDEZ, V.: 'Gigahertz clocked point-to-point and multi-user quantum key distribution systems', PhD Thesis, Heriot-Watt University, UK, 2006

16  HADFIELD R. H., HABIF J.L., SCHLAFER J. S., SCHWALL R. E., NAM S. W. 'Quantum Key Distribution at 1550 nm with twin superconducting single-photon detectors', Applied Physics Letters, 2006, 89, (24), pp 241129—1-241129—3

17  ROSFJORD K. M., YANG J. M., DAULER E.A., KERMAN A. J., ANANT V., VORONOV B. M., GOL'TSMAN G. N., BERGGREN K. K. 'Nanowire single-photon detector with integrated optical cavity and anti-reflection coating' Optics Express, 2006, 14, (2), pp 527-534

**Authors' affiliations:**

R.J. Collins, V. Fernandez and G.S. Buller (School of Engineering and Physical Sciences, David Brewster Building, Heriot-Watt University, Edinburgh, EH14 4AS, United Kingdom and SUPA, The Scottish Universities Physics Alliance, David Brewster Building, Heriot-Watt University, Edinburgh, EH14 4AS, United Kingdom. Email: rjc2@hw.ac.uk)

R.H. Hadfield and S. Nam (National Institute of Standards and Technology, 325 Broadway, Boulder, Colorado 80305, United States of America)

**Figure captions:**

Fig. 1  The quantum key distribution test bed used to measure the performance of the superconducting single photon detector (SSPD).  The initial data-stream to drive the high-speed vertical cavity surface emitting lasers (VCSELs) was a non-return to zero (NRZ) signal from a pulse pattern generator (PPG).  The telecommunications fibre quantum channel was simulated by using additional attenuation in Alice.

Fig. 2  Quantum Bit Error Rate (QBER) and Net Bit Rate (NBR, secret key transmission rate) for the quantum key distribution test bed using the superconducting single photon detector (SSPD).

◊ QBER measured by superconducting single photon detector
■ Net Bit Rate measured by superconducting single photon detector

Figure 1

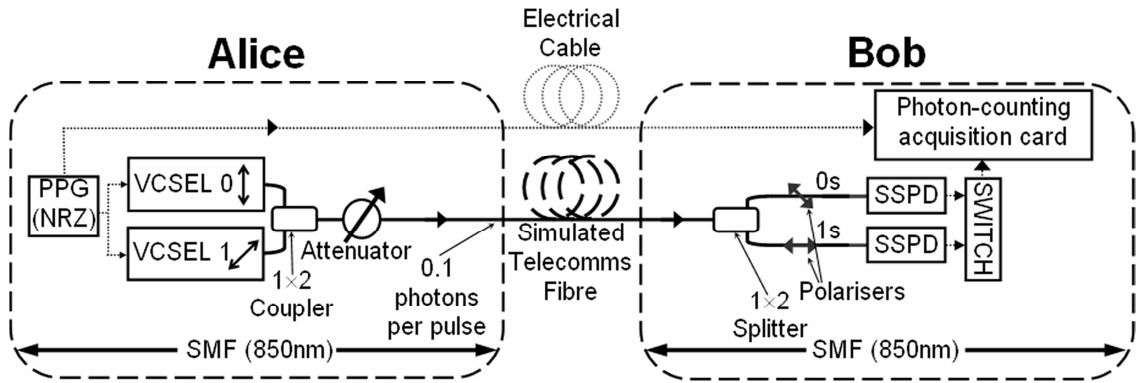

Figure 2

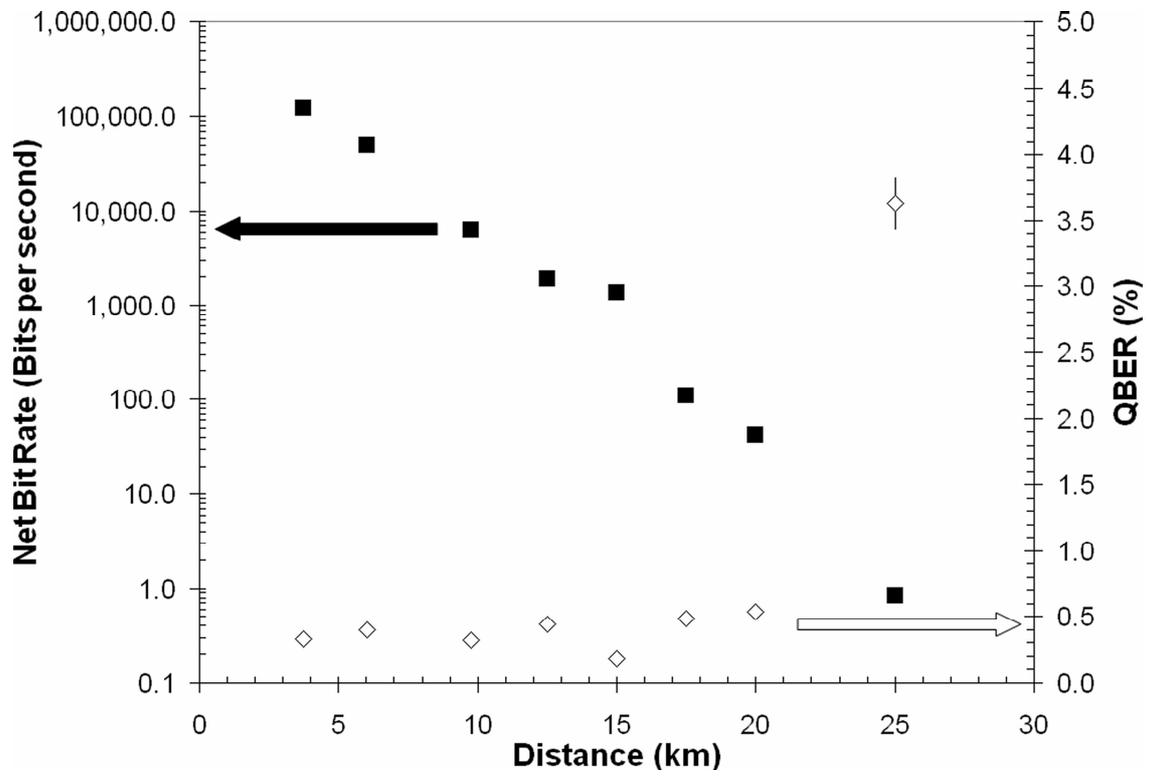